# Demand Side Management in Smart Grids using a Repeated Game Framework

Linqi Song, Yuanzhang Xiao and Mihaela van der Schaar, *Fellow, IEEE*

***Abstract*—**Demand side management (DSM) is a key solution for reducing the peak-time power consumption in smart grids. To provide incentives for consumers to shift their consumption to off-peak times, the utility company charges consumers differential pricing for using power at different times of the day. Consumers take into account these differential prices when deciding when and how much power to consume daily. Importantly, while consumers enjoy lower billing costs when shifting their power usage to off-peak times, they also incur discomfort costs due to the altering of their power consumption patterns. Existing works propose *stationary* strategies for the myopic consumers to minimize their short-term billing and discomfort costs. In contrast, we model the interaction emerging among self-interested, foresighted consumers as a repeated energy scheduling game and prove that the stationary strategies are suboptimal in terms of long-term total billing and discomfort costs. Subsequently, we propose a novel framework for determining optimal *nonstationary* DSM strategies, in which consumers can choose different daily power consumption patterns depending on their preferences, routines, and needs. As a direct consequence of the nonstationary DSM policy, different subsets of consumers are allowed to use power in peak times at a low price. The subset of consumers that are selected daily to have their joint discomfort and billing costs minimized is determined based on the consumers' power consumption preferences as well as on the past history of which consumers have shifted their usage previously. Importantly, we show that the proposed strategies are incentive-compatible. Simulations confirm that, given the same peak-to-average ratio, the proposed strategy can reduce the total cost (billing and discomfort costs) by up to 50% compared to existing DSM strategies.

***Keywords—Smart Grids; Demand Side Management; Critical Peak Pricing; Consumer Discomfort; Repeated Games; Incentive Design.***

## I. INTRODUCTION

Smart grids aim to provide a more reliable, environmentally friendly and economically efficient power system [1][2]. The utility company sells electricity to consumers, who are equipped with smart meters. Smart meters exchange information between consumers and the utility company, and schedule the household energy consumption for consumers. The information gathered through smart meters can be used by the utility company to adjust the electricity prices.

*Demand Side Management (DSM)* is a key mechanism to make smart grids more efficient and cost-effective [1][2]. DSM refers to the programs adopted by utility companies to directly or

L. Song, Y. Xiao and M. van der Schaar are with Department of Electrical Engineering, UCLA. Email: songlinqi@ucla.edu, {yxiao, mihaela}@ee.ucla.edu.

indirectly influence the consumers' power consumption behavior in order to reduce the *Peak-to-Average Ratio (PAR)* of the total load in the smart grid system. A higher PAR results in much higher operation costs and possibly outages of the system. DSM aims to incentivize consumers to shift their peak-time power consumption to off-peak times, thereby resulting in significant PAR reductions in the power system.

### *A. Related Work*

*Direct Load Control (DLC)* and *Smart Pricing (SP)* are two popular existing approaches for implementing DSM. DLC refers to the program in which the utility company can remotely manage a fraction of consumers' appliances to shift their peak-time power usage to off-peak times [3]. Alternatively, SP [6]-[18] provides an economic incentive for consumers to voluntarily manage their power usage. Examples are *Real-Time Pricing (RTP)* [6], *Time-Of-Use Pricing (TOU)* [14], *Critical Peak Pricing (CPP)* [16]-[18], etc. However, the above works [3][6]-[8][16]-[18] do not consider the consumers' discomfort costs which is induced by altering their power consumption patterns.

Some recent works considered consumers' discomfort costs [4][5][9]-[15][20][23] and aimed to jointly minimize the consumers' billing and discomfort costs (referred to subsequently as the total cost). These works can be classified into two categories, depending on the deployed consumer model. The first category assumed that the consumers are price-taking (i.e., they do not consider how their consumption will affect the prices). By assuming that consumers are price-takers, the decision making of a *single foresighted* consumer is formulated as a stochastic control problem aiming to minimize its long-term total cost in [9]-[11]. Alternatively, in [12][13], *multiple myopic* consumers aim to minimize their current total costs and their decisions are formulated as static optimization problems among cooperative users for which distributed algorithms are proposed to find the optimal prices.

The second category assumed that the consumers are *myopic* and price-anticipating (i.e., they take into account how their consumption will affect the prices). In this case, each consumer's power usage affects the other consumers' billing costs. These works [6]-[8][15] modeled the interactions emerging among myopic consumers as one-shot games and studied the *Nash*

TABLE I. COMPARISON WITH EXISTING WORKS

| | **DLC or SP** | **Price-taking or price-anticipating consumers** | **Myopic or foresighted consumers** | **Single consumer or multiple consumers** | **Model** | **Consumer Discomfort** |
|---|---|---|---|---|---|---|
| [3] | DLC | - | Myopic | Multiple | Optimization | No |
| [4][5] | DLC | - | Myopic | Multiple | Optimization | Yes |
| [6]-[8] | SP | Price-anticipating | Myopic | Multiple | One-shot game | No |
| [9]-[11] | SP | Price-taking | Foresighted | Single | Stochastic control | Yes |
| [12][13] | SP | Price-taking | Myopic | Multiple | Optimization | Yes |
| [14] | SP | Price-taking | Myopic | Single | One-shot game | Yes |
| [15] | SP | Price-anticipating | Myopic | Multiple | One-shot game | Yes |
| [16] | SP | Price-taking | Myopic | Multiple | Optimization | No |
| Our Work | SP | Price-anticipating | Foresighted | Multiple | Repeated game | Yes |

*equilibrium (NE)* of the emerging game. In this paper, we also model the consumers as price-anticipating. However, in our model, the consumers interact with each other repeatedly and are foresighted, thereby engaging in a repeated game.

It is well-known that in one-shot games with myopic price-anticipating consumers, the system performance (i.e., the total cost) at the equilibrium can be much worse than the optimal performance achieved by myopic price-taking consumers [25]. One way to achieve the same performance for myopic price-anticipating consumers as in the case of myopic price-taking consumers is to model the system as a repeated game, and to incentivize the price-anticipating consumers to choose (in equilibrium) the optimal consumption adopted by myopic price-taking consumers [24]. Importantly, in this paper we go one step further, and adopt a novel repeated game framework which can significantly outperform the optimal performance obtained by the myopic price-taking consumers [12][13].

All the existing works considering multiple consumers [3]-[8][12]-[16] assumed that the consumers are myopic and minimize their current costs. The optimal DSM strategies in these works are *stationary*, i.e., all consumers adopt fixed daily/weekly power consumption patterns as long as the system parameters (e.g., the consumers' desired power consumption patterns) do not change. However, as we will show later in the paper, the stationary DSM strategies are suboptimal in terms of the long-term total cost. To minimize the total cost, *some* consumers are required to shift their peak-time power usage to the off-peak times while the remaining consumers can use energy when desired. By deploying this optimal strategy, the consumers who shift their peak-time consumption incur discomfort costs, but this leads to a reduction of the peak-time price and of the billing cost of *all* the consumers. Importantly, our proposed nonstationary DSM strategy can achieve the *optimal* total cost while ensuring fairness among consumers by recommending different subsets of consumers (referred to as the *active set*) to shift their peak-time consumption each day. The active set is determined by the consumers' preferences and the past selection of active sets.

A detailed comparison of our work and existing works is highlighted in Table I.

### *B. Our Contributions*

In this paper, since the consumers stay in the system for a long time and interact with each other repeatedly, we formulate the consumers' interactions as a repeated game. The repeated nature of the interaction provides incentives for price-anticipating consumers to cooperate (as shown in [24]). Although the proposed framework can improve the performance of the stationary DSM strategies discussed for any SP scheme, we focus on the CPP scheme, which has been widely used for residential consumers and is shown to work well in practical scenarios [17]-[19]. CPP defines peak days in a year or peak times in a day, and charges higher prices during these peak hours if CPP events, such as system load warning, extreme weather conditions, and system emergencies, occur [18].

Based on the repeated game model, we propose an optimal nonstationary[1] DSM mechanism that minimizes the total cost and outperforms the optimal stationary DSM strategy. In addition, the proposed strategy is *Incentive-Compatible (IC)*, namely the self-interested consumers will find it in their self-interest to follow the recommended strategy. Each day, the DSM strategy selects an active set of consumers based on their preferences of whether or when to shift and on the past history of consumption pattern shifts. These consumers sacrifice their current discomfort costs to minimize the total billing cost. In return, they will enjoy in the future lower billing costs without incurring discomfort costs when other consumers are chosen in the active set. In this way, the proposed strategy minimizes the long-term total cost while ensuring fairness among the consumers.

In summary, the main contributions of our work are as follows:

[1] Recall that in a stationary DSM mechanism, the consumers adopt fixed daily/weekly power consumption patterns as long as the system parameters do not change. In contrast, in a nonstationary DSM mechanism, the consumers may adopt different daily/weekly power consumption patterns (e.g., a consumer may shift its peak-time consumption today but not shift tomorrow) even if the system parameters remain the same.

- *A Repeated Game Framework:* A repeated game framework is proposed to model the interactions among foresighted price-anticipating consumers over time.
- *Joint Billing and Discomfort Costs Minimization:* The proposed DSM mechanism considers not only the consumers' billing costs but also their discomfort costs.
- *Optimal Nonstationary DSM mechanism:* We analytically prove that in the energy scheduling game with discomfort costs, nonstationary DSM mechanisms can outperform stationary mechanisms, and propose an optimal nonstationary DSM mechanism that can be easily implemented.
- *IC strategies:* The DSM mechanism is IC, meaning that the self-interested consumers have no incentive to deviate from the recommended strategy.
- *Consumer heterogeneity:* Our framework can model different types of consumers with different discomfort costs. Moreover, different consumers may have different preferences on how and when their consumption patterns are shifted. We study the impact of different types of consumers on the performance of the proposed mechanism.

The rest of this paper is organized as follows. Section II models the repeated energy scheduling game and formulates the DSM mechanism design problem. Section III formally introduces the proposed algorithm for constructing the DSM strategy and discusses how to implement the proposed algorithm in the smart grid system. Section IV provides simulation results to validate the performance of the proposed algorithm. Section V concludes the paper.

## II. System Model

### *A. Energy Scheduling Game*

A smart grid system consists of a utility company and multiple consumers, as shown in Fig.1. The DSM is implemented through smart meters on the consumer side and a DSM center in the utility company [2].

We denote the set of consumers by $\mathcal{N} = \{1,2,\ldots,N\}$ . Time is divided into periods $t = 0,1,2,\cdots$. We assume that each period is divided into $H \in \mathbb{N}_+$ time slots with equal length and denote the set of time slots by $\mathcal{H} = \{1,2,\ldots,H\}$. Note that we use "period" here to denote each stage of the interaction among consumers and use "time slot" to denote the discrete time to schedule power usage within a period. In this paper, we consider a period to be one day as in [6]-[9][12]-[16], and each slot can be one or multiple hours (e.g., $H = 24$, $H = 12$).

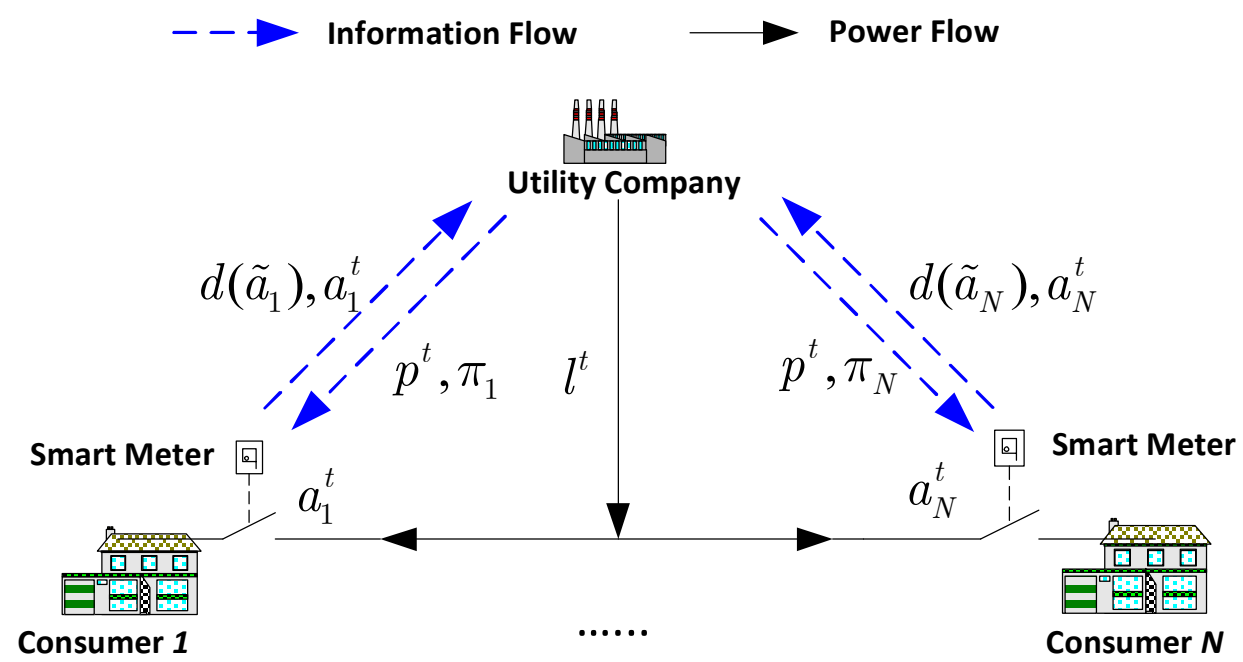

Fig. 1. Smart Grid System Model.

For each consumer $i \in \mathcal{N}$ in period $t$, its action is the power consumption pattern in that period, which is the vector of power consumption at each time slot and denoted by $a_i^t = (a_{i,1}^t, \ldots, a_{i,H}^t)$, with $a_{i,h}^t \in \mathcal{A}_i$ being the power consumption and $\mathcal{A}_i$ being the set of power consumption at each time slot.

The power consumption $a_{i,h}^t$ at time slot $h$ consists of non-shiftable load and shiftable load. The non-shiftable load, such as lighting, cooking, watching TV, is not controllable by smart meters, while the shiftable load, such as dish and clothes washing, heating and cooling systems, can be controlled by the smart meters [6]-[8]. We denote the non-shiftable and shiftable loads at time slot $h \in \mathcal{H}$ by $b_{i,h}^t \geq 0$ and $s_{i,h}^t \geq 0$, respectively. Thus, consumer $i$'s power consumption satisfies $a_{i,h}^t = b_{i,h}^t + s_{i,h}^t$ and $a_{i,h}^t \geq b_{i,h}^t$. Note that here the shiftable load is not only limited to the deferrable load, but also includes "smart thermostats" aimed at controlling heating/cooling systems, which can shift the energy usage ahead of time and have been shown [19] to significantly reduce the peak load in practice, etc.

We denote by $\sum_{h=1}^{H} a_{i,h}^t = A_i$ the daily total power consumption for residential consumer $i$, where $A_i$ is either a constant or slowly varying as in [6]-[8][13]-[16]. We denote by $\boldsymbol{a}^t = (a_1^t, a_2^t, \cdots a_N^t) \in \mathcal{A}$ the power consumption profile of all consumers, where $\mathcal{A} = \times_{i=1}^{N} \mathcal{A}_i^H$.

The *total load* at time slot $h$, denoted by $l_h^t = \sum_{i=1}^{N} a_{i,h}^t$, is the sum of all consumers' power consumption.

We assume that the desired power consumption pattern in each period for consumer $i$ is $\overline{a}_i = [\overline{a}_{i,1}, \overline{a}_{i,2}, \cdots \overline{a}_{i,H}] \in \mathcal{A}_i^H$, which refers to its preferred daily power consumption pattern [22]. The corresponding desired power consumption profile of all consumers and total load are denoted by $\overline{\boldsymbol{a}} = [\overline{a}_1, \overline{a}_2, \cdots \overline{a}_N] \in \mathcal{A}$ and $\overline{l}_h = \sum_{i=1}^{N} \overline{a}_{i,h}$. Define $\overline{h} = \arg\max_{h \in \mathcal{H}} \overline{l}_h$ as the peak time of the day, the length of which changes according to how $H$ is set. Empirical studies show that, compared with industrial and commercial consumers, residential consumers have very similar peak-time shiftable loads [19], implying that $\overline{a}_{i,\overline{h}} - b_{i,\overline{h}}^t = \overline{s}_{\overline{h}}$, for each consumer $i$.

*Remark:* First, different consumers may have different desired power consumption patterns $\overline{a}_i$. For example, most consumers may have their highest power consumption in the evening, while some consumers may have highest power consumption in the morning. Second, each consumer can have its individual peak consumption in hours other than $\overline{h}$; $\overline{h}$ is determined based on the aggregate power consumption of all consumers, instead of the power consumption of individual consumers. Third, in the model currently presented, we let the desired daily power consumption pattern $\overline{a}_i$ to be the same for each day. One may argue that the desired daily power consumption should be different in weekdays and in weekends. In this case, we can easily extend the model such that one period is one week and $\overline{a}_i$ is the desired weekly power consumption pattern, where the desired daily power consumption patterns are different each day.

The cost of a consumer consists of billing and discomfort costs. The billing cost is the power consumed multiplied by the unit price and the discomfort cost is the consumer's discomfort caused by rescheduling its daily power consumption from its desired power consumption pattern. We denote $c_i : \mathcal{A} \mapsto \mathbb{R}$ to be the cost function of consumer $i$:

$$c_i(\boldsymbol{a}^t) = \sum_{h=1}^{H} p_h(\boldsymbol{a}^t) a_{i,h}^t + d_i(a_i^t), \tag{1}$$

where $p_h : \mathcal{A} \mapsto \mathbb{R}_+$ denotes the price at time slot $h$, $\sum_{h=1}^{H} p_h(\boldsymbol{a}^t) a_{i,h}^t$ denotes the daily billing cost, $d_i : \mathcal{A}_i^H \mapsto \mathbb{R}$ denotes the discomfort cost of consumer $i$. Without loss of generality, the total cost is the sum, instead of the weighted sum, of billing and discomfort costs, because any weight the consumer puts on the discomfort cost can be absorbed in the expression of the discomfort cost function $d_i$. We will discuss later the price $p_h$ and discomfort cost $d_i$ in detail.

To model the interactions among consumers, we first formalize the one-shot energy scheduling game $G = \{\mathcal{N}, \{\mathcal{A}_i\}_{i=1}^N, \{c_i\}_{i=1}^N\}$, where $\mathcal{N}$, $\{\mathcal{A}_i\}_{i=1}^N$ and $\{c_i\}_{i=1}^N$ denote the set of consumers, sets of actions and cost functions for each consumer, respectively.

Next, we formalize the repeated game model. In each period $t$, consumer $i$ determines its power consumption pattern $a_i^t$ based on its history, which is a collection of all its past power consumption patterns and the past prices made public to the consumers. The history of consumer $i$ up to period $t$ can be written as: $\eta_i^t = \{a_i^0, p^0, a_i^1, p^1, ..., a_i^{t-1}, p^{t-1}\} \in (\mathcal{A}_i^H \times \mathcal{P}^H)^t$ for $t > 0$ and the initial history is defined as $\eta_i^0 = \varnothing$. The corresponding *public history* is defined as $\eta^t = \{p^0, p^1, ..., p^{t-1}\} \in (\mathcal{P}^H)^t$ for $t > 0$ and the initial history is defined as $\eta^0 = \varnothing$. The *public strategy* of consumer $i$ is defined as a mapping from public history to current action, denoted by $\pi_i : \bigcup_{t=0}^{\infty} (\mathcal{P}^H)^t \mapsto \mathcal{A}_i$, where $(\mathcal{P}^H)^0 = \varnothing$ [24]. Due to realization equivalence principle [24, Lemma 7.1.2], we lose nothing, in terms of the achievable operating points, by restricting to public strategies, compared to strategies using the entire history.

Given the strategy profile of all consumers, denoted by $\boldsymbol{\pi} = (\pi_1, \pi_2, \cdots \pi_N)$, consumer $i$'s average long-term cost is discounted by a factor $\delta$:

$$C_i(\boldsymbol{\pi}) = (1 - \delta) \sum_{t=0}^{\infty} \delta^t c_i(\boldsymbol{\pi}(\eta^t)), \tag{2}$$

where $c_i(\boldsymbol{\pi}(\eta^t))$ is the cost of consumer $i$ in period $t$. The discount factor $\delta \in [0,1)$ represents how consumers discount their monetary costs in the future, e.g. due to interest rates/inflation. A smaller $\delta$ means that the consumers discount their future costs more. Since the interest rate/inflation is usually the same for all the consumers, we assume $\delta$ to be equal for all consumers as in [24][25]. The corresponding long-term discounted discomfort cost is denoted by $D_i(\boldsymbol{\pi})$.

Hence, the infinitely repeated game model can be written as $RG = \{\mathcal{N}, \bigcup_{t=0}^{\infty}(\mathcal{P}^H)^t, \{\pi_i\}_{i=1}^N, \{C_i(\boldsymbol{\pi})\}_{i=1}^N\}$, where $\mathcal{N}$, $\bigcup_{t=0}^{\infty}(\mathcal{P}^H)^t$, $\{\pi_i\}_{i=1}^N$ and $\{C_i(\boldsymbol{\pi})\}_{i=1}^N$ denote the set of consumers, set of public histories, sets of strategies and sets of cost functions, respectively.

*B. Critical Peak Pricing Scheme*

Recall that in Critical Peak Pricing (CPP) scheme, the utility company charges a higher price in the critical peak time when CPP events occur (such as system load warning, extreme weather conditions, and system emergencies), aiming at reducing the peak-time load of the system [16]-[18]. The maximum number of critical days and of critical hours within a day is predefined.

We model the CPP pricing scheme with a single critical peak time and only consider the CPP events triggered by the total load in the system. The time-varying price function $p_h(\boldsymbol{a}^t)$ is defined as:

$$p_h(\boldsymbol{a}^t) = p_h(l_h^t) = \begin{cases} p_{Lo}, & 0 \le l_h^t \le l_{th} \\ p_{Hi}, & l_h^t > l_{th} \end{cases}, \tag{3}$$

where $p_{Hi} > p_{Lo}$ are the peak price and off-peak price of the pricing model and $l_{th}$ is the threshold of the total load. When $l_h^t \le l_{th}$, the higher price will not be triggered and $p_{Lo}$ will be adopted. When $l_h^t > l_{th}$, the CPP event occurs and the higher price $p_{Hi}$ will be adopted. To provide incentives for consumers to shift their peak-time loads, the threshold $l_{th}$ is set to be below the peak load and above the off-peak load, namely,

$$\overline{l}_{\overline{h}} > l_{th}, \quad \overline{l}_{h'} \le l_{th}, \quad \forall h' \in \mathcal{H} \backslash \{\overline{h}\}. \tag{4}$$

Given peak load reduction goal, namely $l_{th}$, we further set $m = \left(\overline{l}_{\overline{h}} - l_{th}\right) / \overline{s}_{\overline{h}}$, where $m \in \mathbb{N}$ is the smallest number of consumers needed to shift their peak-time consumption such that the peak-time price is low. We denote the prices within a day by $\boldsymbol{p} = (p_1, p_2, \cdots p_H)$.

*C. Consumer Discomfort Cost*

We use a discomfort cost function to model the consumers' discomfort from rescheduling their power consumption patterns. Many papers define the discomfort cost function based on the imposed change in their consumption patterns, i.e., the "distance" between consumer's desired demand and actual consumption [14][20]-[23]. As in [14][20], we use a linear weighted function to model the discomfort cost:

$$d_i(a_i^t) = \begin{cases} \sum_{h=1}^{H} k_{i,h}(|\, a_{i,h}^t - \overline{a}_{i,h} \,|) + \omega_i, & a_i^t \neq \overline{a}_i \\ 0, & a_i^t = \overline{a}_i \end{cases}, \tag{5}$$

where $k_{i,h}, \omega_i \in \mathbb{R}_+$ are parameters of the discomfort cost function. Consumer $i$'s discomfort cost $d_i(a_i)$ is an increasing function of the 'distance' between the rescheduled and the desired power consumption patterns.

The above discomfort cost function is able to capture several important consumer preferences in terms of consumption pattern shifting. First, $\omega_i$ captures the consumer's *willingness* to shift: a consumer with a large $\omega_i$ is less willing to shift. Second, $k_{i,h}$ captures the consumer's preference on *how to shift*. A larger $k_{i,\overline{h}}$ indicates that the consumer does not want to reduce its peak-time consumption; a larger $k_{i,h}$ for $h \neq \overline{h}$ indicates that the consumer does not want to shift its peak-time consumption to time slot *h*.

We are interested in two costs that consumer $i$ can achieve: consumer $i$'s minimum cost achievable by any power consumption profile, denoted by $\underline{c}_i = \min_{\boldsymbol{a} \in \mathcal{A}} c_i(\boldsymbol{a}) = p_{Lo} A_i$, and consumer $i$'s minimum cost achievable by any power consumption profile in which consumer $i$ shifts all its peak-time shiftable load, denoted by $\tilde{c}_i = \min_{\boldsymbol{a} \in \mathcal{A},\, a_{i,\overline{h}} = b_{i,\overline{h}}} c_i(\boldsymbol{a}) = p_{Lo} A_i + d_i(\tilde{a}_i)$, where $\tilde{a}_i = \operatorname*{argmin}_{a_i \in \mathcal{A}_i^H,\, a_{i,\overline{h}} = b_{i,\overline{h}}} d_i(a_i)$. Clearly, we always have $\overline{c}_i > \tilde{c}_i$, because $\tilde{c}_i$ is the minimum cost achievable when constrained to consumer $i$'s certain power consumption patterns (which cause discomfort costs). The cost $\underline{c}_i$ is important because it is the minimum cost that consumer $i$ could possibly achieve. The cost $\tilde{c}_i$ is also important because it represents the minimum cost that consumer $i$ could possibly achieve when it shifts all of its peak-time consumption (i.e. when it is selected to be in the active set).

Based on the relationship between billing and discomfort costs, we can classify the consumers into three classes.

- *Consumers with low discomfort costs:* This class includes consumers whose discomfort costs are low compared with $p_{Hi} - p_{Lo}$. In this case, the increase in the price matters much more to consumers than the discomfort costs. For these consumers, the discomfort costs can be reasonably ignored and only the billing costs need to be considered.

- *Consumers with medium discomfort costs:* This class includes consumers who have medium discomfort costs with respect to $p_{Hi} - p_{Lo}$. In this case, both billing and discomfort costs need to be considered. We say a consumer has medium discomfort cost if

$$\frac{(p_{Hi} - p_{Lo})\overline{l}_{\overline{h}}}{m} > d_i(\tilde{a}_i) \ and \ \omega_i > [(p_{Hi} - p_{Lo})]\overline{a}_{i,\overline{h}}. \tag{6}$$

The first inequality implies that the discomfort cost is not too large, such that the consumers are willing to shift their peak-time consumption if by doing so their billing costs can be greatly reduced. The second inequality implies that the discomfort cost is not too small, such that each consumer does care about its own discomfort cost and is not willing to shift its peak-time consumption every day.

- *Consumers with high discomfort costs:* This class includes consumers who have high discomfort costs parameters compared with $p_{Hi} - p_{Lo}$. In this case, discomfort costs are high when altering their power consumption patterns and hence they should be recommended not to change their power consumption patterns.

In a practical system, the above three classes of consumers coexist. However, since the recommendations to consumers with low and high discomfort costs can be determined based on their cost characteristics (namely, consumers with low discomfort costs always shift and choose $\tilde{a}_i$, while consumers with high discomfort costs cannot shift their power consumption pattern and choose $\overline{a}_i$), we only need to consider the DSM strategy for consumers with medium discomfort costs. Hence, in Section III, we focus on deriving the optimal DSM strategy for consumers with medium discomfort costs. The optimal DSM strategy derived for consumers with medium discomfort costs, along with the predetermined recommended strategies for consumers with high and low discomfort costs, constitutes the DSM strategy used in the practical system with three classes of consumers.

*D.Problem Formulation*

In this subsection, we consider the system with self-interested consumers and formulate the optimal IC DSM mechanism design problem.

The designer is the benevolent utility company that aims at minimizing the total cost (maximizing the social welfare) in the smart grid system. However, except for the total cost, maintaining fairness among all the consumers is essential [19]. Hence, the mechanism will ensure that the average discomfort cost of consumer $i$ is no greater than a maximal value $D_{i,\max}$. Therefore, the *DSM mechanism Design Problem (DDP)* can be formulated as

$$\begin{aligned}
\text{(DDP): } \underset{\pi}{\text{minimize}} \quad & \sum\nolimits_{i\in\mathcal{N}} C_i(\boldsymbol{\pi}) \\
\text{subject to} \quad & a_{i,h}^t \geq b_{i,h}^t, \forall t \in \mathcal{T} \\
& A_i = \sum\nolimits_{h=1}^{H} a_{i,h}^t, \forall t \in \mathcal{T} \ . \\
& D_i(\boldsymbol{\pi}) \leq D_{i,\max}, \forall i \in \mathcal{N} \\
& \boldsymbol{\pi} \ is\ IC
\end{aligned}$$

The utility company will solve this problem, then recommend the consumers with the optimal solution $\boldsymbol{\pi}^*$.

The abovementioned energy scheduling game model can be extended or revised in different ways to accommodate various systems. First, the CPP scheme may be only applied in the summer and winter seasons, when the system load is high. In this case, the DSM mechanism can update the system parameters and begin to solve the DDP problem again, at the beginning of a new season. Second, the daily power consumption $A_i$ for a specific consumer $i$ can vary for different days during a week, due to their different weekly routines, e.g. they do the laundry on Fridays etc.

In this case, we can set the period in the model to be a week, instead of a day, and set the daily total power consumption and desired power consumption pattern as $\sum_{h=1}^{H} a_{i,h}^{t} = A_i^{t \bmod 7}$ and $\overline{a}_i^t = \overline{a}_i^{t \bmod 7}$, where $\{A_i^\tau\}_{\tau=0}^{6}$ and $\{\overline{a}_i^\tau\}_{\tau=0}^{6}$ are the daily power consumption and desired power consumption patterns of a week and are different for different days of a week.

## III. Optimal Strategies

In this section, we solve the DDP problem defined in the previous section. We first discuss a benchmark case - the performance of the one-shot energy scheduling game. Subsequently, we characterize the Pareto efficient operating points of the repeated energy scheduling game and propose our nonstationary algorithm that achieves the optimal solution of the DDP problem. Finally, we describe the implementation of the proposed DSM in the smart grid system.

### *A. Benchmark – The One-shot Energy Scheduling Game*

In the unique NE of the one-shot energy scheduling game, the consumer chooses its desired power consumption pattern. We state this formally in the following theorem.

*Theorem 1 (Nash Equilibrium of the One-shot Game)*: The one-shot energy scheduling game has a unique *NE*, in which each consumer chooses its desired power usage as

$$a_i^* = \overline{a}_i, \forall i \in \mathcal{N} . \tag{7}$$

*Proof*: The idea of the proof is to show that $\overline{a}_i$ is the dominant strategy for consumer $i$. The complete proof is given in Appendix A.□

The result of Theorem 1 is the well-known "tragedy of commons". Each consumer aims to minimize its individual billing and discomfort costs and thus, it will myopically find in its self-interest to minimize its individual cost by sticking to its desired power consumption pattern. This results in a high price and low social welfare. According to Theorem 1, the total cost at the unique NE is $\sum_{i=1}^{N} C_{i,NE}$, where $C_{i,NE} = c_i(\overline{\boldsymbol{a}})$. In the following section we will quantify the inefficiency of the NE and propose a novel DSM mechanism that can achieve the optimal social welfare.

*B. Pareto-optimal Region of the Repeated Energy Scheduling Game*

We formally characterize the achievable operating points of the repeated energy scheduling game. It is well known that as long as $\delta$ is sufficiently close to 1, the achievable region of the repeated energy scheduling game is the convex hull of the one-shot energy scheduling game [24, Lemma 3.7.1]. Hence, we can write the achievable region of repeated energy scheduling game as $\mathcal{C} = conv\{\boldsymbol{c}(\boldsymbol{a}) \mid \boldsymbol{a} \in \mathcal{A}, \sum_{h=1}^{H} a_{i,h} = A_i, a_{i,h} \geq b_{i,h}\}$, where $conv\{X\}$ is the convex hull of $X$ and $\boldsymbol{c}(\boldsymbol{a}) = (c_1(\boldsymbol{a}), c_2(\boldsymbol{a}), \cdots c_N(\boldsymbol{a}))$ is the cost profile of the consumers. We will prove that the Pareto-optimal region of the repeated energy scheduling game (i.e., the Pareto boundary of the set of achievable cost profiles), denoted by $\mathcal{B}$, is a face of $\mathcal{C}$ that is a polytope with dimension $N-1$. This means that we can analytically express the Pareto-optimal region. Theorem 2 formally characterizes the Pareto-optimal region $\mathcal{B}$ analytically.

*Theorem 2*: The Pareto-optimal region of the repeated energy scheduling game is

$$\mathcal{B} = \{\boldsymbol{C} = (C_1, C_2, \cdots, C_N) \mid \sum_{i=1}^{N} (C_i - \tilde{c}_i) / (\bar{c}_i - \tilde{c}_i) = m, C_i \geq \tilde{c}_i\}. \tag{8}$$

In addition, the stationary DSM strategies can only achieve the extreme points[2] of $\mathcal{B}$.

*Proof*: See Appendix B. □

Theorem 2 does not only characterize the Pareto-optimal region (i.e., part of a hyperplane) of the repeated energy scheduling game, but also proves that by using stationary DSM strategies, we cannot achieve any points on the Pareto-optimal region other than the extreme points. Note, however, that, the extreme points can only be achieved by the action profiles in which $m$ consumers shift their peak consumption and incur cost $\bar{c}_i$, while the other consumers do not shift and incur cost $\tilde{c}_i$. It is clear that the extreme points are not desirable operating points, because an extreme point can be achieved only when a fixed set of $m$ consumers shift their peak-time consumption all the time, which is unfair for these $m$ consumers because they incur high discomfort costs all the time. Hence, the desired operating points lie in the interior of the Pareto-optimal region, which can be achieved by nonstationary strategies in the repeated energy scheduling game according to Theorem 2.

---

[2] An extreme point of a convex set is the point that is not the convex combination of any other points in this set. In our case, since $\mathcal{B}$ is part of a hyperplane, the extreme points will be the vertices of $\mathcal{B}$.

By adding IC constraints and the constraints on the maximum discomfort costs, the *feasible* Pareto-optimal region can be written as

$$\mathcal{B}_{\bar{C}} = \{\boldsymbol{C} = (C_1, C_2, \cdots, C_N) \mid \sum_{i=1}^{N} (C_i - \tilde{c}_i) / (\bar{c}_i - \tilde{c}_i) = m, C_i \geq \tilde{c}_i, C_i \leq \bar{C}_i\}, \quad (9)$$

where $\bar{C}_i = \min\{C_{i,\max}, C_{i,NE}\}$, $C_{i,\max} = \tilde{c}_i + D_{i,\max}$.

*C. Nonstationary DSM Mechanism*

Given the Pareto-optimal region, we can then reformulate the *DDP* problem as a linear programming problem:

$$\begin{array}{ll} \text{minimize} & \sum_{i \in \mathcal{N}} C_i \\ \text{subject to} & \boldsymbol{C} \in \mathcal{B}_{\bar{C}} \end{array}. \quad (10)$$

The solution of (10) is an extreme point of $\mathcal{B}_{\bar{C}}$, denoted by $\boldsymbol{C}^* = (C_1^*, C_2^*, \cdots C_N^*)$.

*Theorem 3*: The feasible Pareto-optimal region $\mathcal{B}_{\bar{C}}$ is achievable if the discount factor $\delta$ satisfies

$$\delta \geq 1 - \frac{1}{N - m + 1}. \quad (11)$$

*Proof*: See Appendix C. □

Given a desired operating point in $\mathcal{B}_{\bar{C}}$, we can use the *Nonstationary DSM (N-DSM)* algorithm, described in Table II, to construct the DSM strategy. In period $t$, the N-DSM algorithm chooses the active set $I(t) \in \mathcal{I}$ consisting of $m$ out of $N$ consumers to reschedule their power consumption patterns, where $\mathcal{I}$ is the set of all possible index combination that containing $m$ consumers out of $N$. The choice of which $m$ consumers are selected depends on "how far" they are from their target cost and this is measured by index $g_i(t)$ and the $m$ consumers who have the largest $g_i(t)$ will alter their power consumption patterns.

*Theorem 4*: When the discount factor $\delta$ satisfies (11), the N-DSM algorithm is IC and can achieve the optimal operating point $\boldsymbol{C}^* = (C_1^*, C_2^*, \cdots C_N^*)$.

*Proof*: See Appendix D. □

TABLE II. NONSTATIONARY DSM (N-DSM) ALGORITHM

**Input**: Target average cost vector $\boldsymbol{C}^* = (C_1^*, C_2^*, \cdots C_N^*)$ , $t = 0$ .
**Output**: Optimal strategy

Set $g_j(t) = (C_j^* - \tilde{c}_j) / (\overline{c}_j - \tilde{c}_j)$ .
**Repeat**
    **If** $p_{\overline{h}}(\tau) = p_{Hi}, \exists \tau < t$ , then
        Recommend action $\overline{a}_i$ to all consumer $i$ .
    **Else**
        Find the active set $I(t) = \{i_1, i_2, \cdots i_m\}$ of $m$ consumers who have the $m$ largest indices $g_j(t)$ .
        Recommend action $\tilde{a}_i$ to $i \in I(t)$ , and $\overline{a}_i$ to $i \notin I(t)$ .
        Observe consumers' action $a_i$ .
        **If** all consumers follow the recommendation, then
            Update $g_i(t+1) = [g_i(t) - (1-\delta)\mathbf{1}_{\{i \in I(t)\}}] / \delta$ for all $i$ .
            Broadcast $p_h(t) = p_{Lo}$ for all $h$ .
        **Else**
            Broadcast $p_{\overline{h}}(t) = p_{Hi}$ and $p_h(t) = p_{Lo}, h \neq \overline{h}$ .
        **End if**
    **End if**
    $t \leftarrow t+1$
**End Repeat**

Theorem 3 and 4 state that when the discount factor satisfies (11), the optimal nonstationary DSM mechanism can be constructed by the N-DSM algorithm. In practice, the daily interest rate is low enough to make condition (11) be satisfied and the proposed DSM mechanism can accommodate a large number of consumers.

### *D. DSM Implementation*

In this subsection, we describe how to implement the DSM mechanism in the smart grid system based on the proposed N-DSM algorithm. The public information, such as price information and total load, is known to every consumer and the private information of consumer $i$, such as discomfort cost, power consumption, is only known to its own. There are two phases in implementing the DSM mechanism: the *initialization phase* and the *run time phase*. As in [2], the utility company recommends the strategy one day in advance.[3] Consumers can schedule their power usage and exchange information with utility company through smart meters for the upcoming day so as to minimize their billing and discomfort costs based on the recommendation.

[3] The DSM can also be implemented in real-*time* or *T-day-ahead*, as long as there is enough time for the consumers to respond to the recommendations.

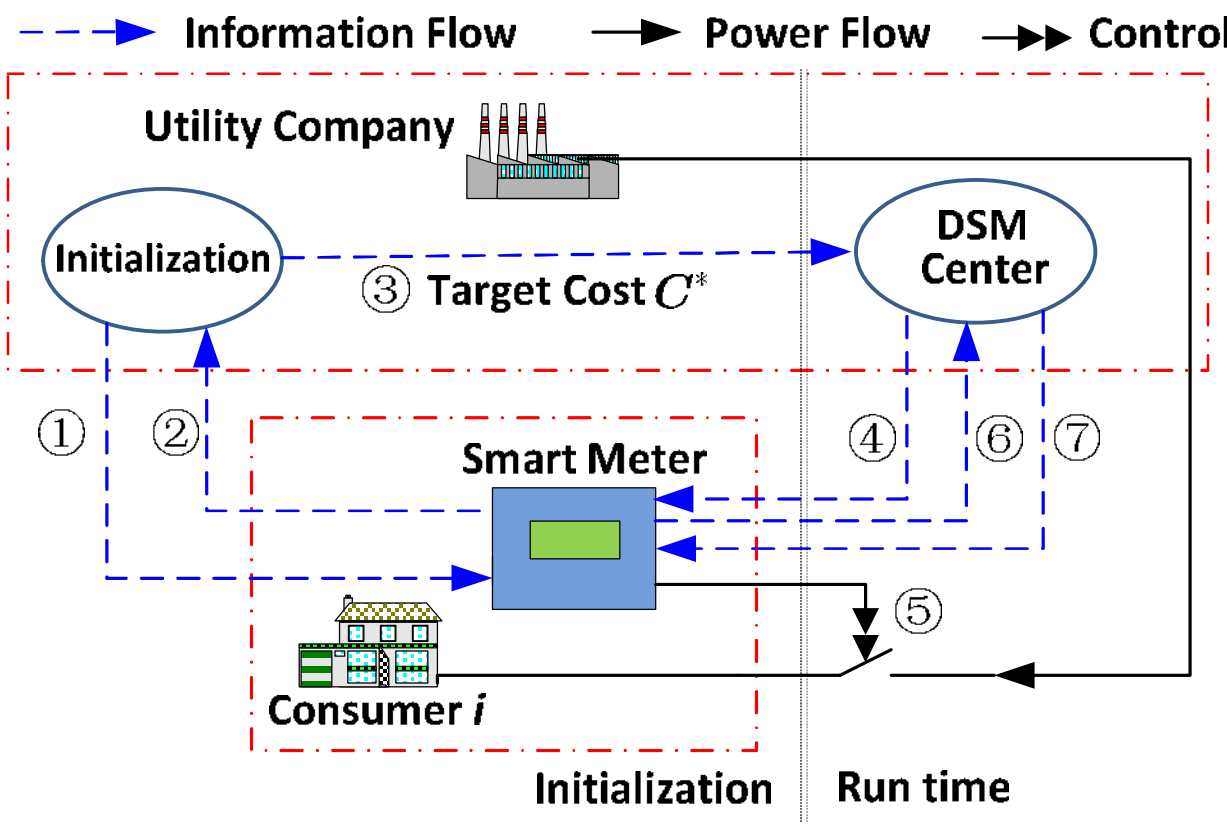

Fig.2. The Implementation of DSM.

TABLE III. INFORMATION EXCHANGE OF IMPLEMENTATION

| | | | |
|---|---|---|---|
| Public information (known to all consumers) | | | $p, p_{Hi}, p_{Lo}, l_{th}(m), l_h^t$ |
| Consumer $i$ 's private information | | | $d_i(a_i^t), a_i^t$ |
| Information exchange | Utility company to consumer $i$ | Initialization | ①: $p_{Hi}, p_{Lo}, l_{th}, \overline{l}_h, \boldsymbol{C}^*$ |
| | | Run time | ④: $\pi_i$, ⑦: $p$ |
| | Consumer $i$ to utility company | Initialization | ②: $d_i(\tilde{a}_i)(or\ \overline{c}_i, \tilde{c}_i), \overline{a}_i$ |
| | | Run time | ⑤⑥: $a_i^t$ |

The details of public and private information, the two phases and the corresponding information exchange of the two phases are shown in Fig. 2 and Table III.

*Initialization:* ① The utility company broadcasts the CPP parameters: $p_{Lo}$, $p_{Hi}$ and $l_{th}$. ② Consumer $i$ calculates its cost parameters $\tilde{c}_i$ and $\overline{c}_i$, and sends them to the utility company. ③ The utility company calculates the target cost vector $\boldsymbol{C}^* = (C_1^*, C_2^*, \cdots C_N^*)$ by solving DDP problem, and then sends the target cost vector to all consumers.

*Run time:* In each period: ④ The DSM center calculates the active set $I(t)$ and then makes the recommendation $\pi_i(t)$ to each consumer. ⑤ Each consumer chooses its energy scheduling strategy. Note that the consumers can deviate from the recommended strategy. However, since the strategy is IC, it is in their self-interest to comply with the recommendation. ⑥ The smart meter reports the power consumption patterns $a_i^t$ to DSM center. ⑦ The utility company calculates the price based on the total load and then broadcasts the prices. Go to next period and repeat ④-⑦.

The consumers' discomfort cost functions may vary over time. For example, they may have different desired power consumption patterns in different seasons/months, or they may prefer to shift the peak-time consumption to different off-peak times (i.e., different $k_{i,h}$) due to their change of schedules (e.g., due to a new job), or they simply change their willingness (i.e., different $\omega_i$) to shift. To accommodate for such occasional changes (e.g., once a month), the utility company can run the initialization phase again in order to calculate the new optimal target cost vector, and then run the N-DSM algorithm based on the updated target cost vector.

## IV. Numerical Results

In this section, we compare the performance of our proposed DSM mechanism with those obtained using the one-shot energy scheduling games with myopic price-anticipating consumers [6]-[8][15], joint optimization with myopic price-taking consumers [12][13], as well as stochastic control methods with a single foresighted consumer [9]-[11]. Then we study the impact of different consumer preferences. In addition, we study the system performance when the percentage of shiftable load and the length of the peak time vary.

Throughout this section, we use the following system parameters by default unless we change some of them explicitly. We consider the scenario that $H = 24$ time slots and set the discount factor of the consumers to be $\delta = 0.995$, which ensures that (11) is satisfied. The pricing scheme sets the peak price and off-peak price to be $p_{Hi} = 0.8$ \$/kWh and $p_{Lo} = 0.1$ \$/kWh[4]. According to the utility company's PAR goal (namely the percentage of reduction in peak-time load, written as $\left(\overline{l}_{\overline{h}} - l_{th}\right) / \overline{l}_{\overline{h}}$), the threshold $l_{th}$ in (3) will be set to an appropriate value to control the parameter $m$. We simulate both the scenario with heterogeneous consumers with parameters shown in Table IV and Fig. 3 and the scenario with homogeneous consumers with the same parameters as Type 1 consumers described in Table IV and Fig. 3. In this experiment, the shiftable load of each consumer is set to be 40% of the consumer's total load.

### A. Comparison with Existing Mechanisms

In this subsection we compare our proposed N-DSM algorithm with the existing ones [6][7].

TABLE IV. Parameters of Three Types of Consumers

| | $A_i$ (kWh) | $k_{i,h}(h=1\ to\ 14)$ / $k_{i,h}(h=15\ to\ 24)$ (\$/kWh) | $\omega_i$ (\$) | $D_{i,\max}$ (\$) |
|---|---|---|---|---|
| Type 1 | 10 | 0.2/0.1 | 0.7 | 0.71 |
| Type 2 | 8 | 0.1/0.05 | 1.5 | 0.91 |
| Type 3 | 11 | 0.15/0.1 | 1.2 | 0.95 |

TABLE V. Comparisons of Different Mechanisms

| Works | Algorithm | Strategies |
|---|---|---|
| [6]-[8] [15] | OG-DSM | $\min_{a_i}\{\sum_{h=1}^{H} p_h(\boldsymbol{a})a_{i,h} + d_i(a_i)\}$ |
| [12][13] | JO-DSM | $\min_{\boldsymbol{a}} \sum_{n=1}^{N}[\sum_{h=1}^{H} p_h(\boldsymbol{a})a_{i,h} + d_i(a_i)]$ |
| [9]-[11] | SC-DSM | $\min_{a_i}\{\sum_{h=1}^{H} p_h a_{i,h} + (1-\varepsilon)d_i(a_i)\}$ |
| Our work | N-DSM | $\min_{\boldsymbol{\pi}_i}\{(1-\delta)\sum_{t=0}^{\infty} \delta^t c_i^t(\boldsymbol{\pi}(\eta^t))\}$ |

[4] According to [17][19], the peak price is often at least 6 times higher than the off-peak price.

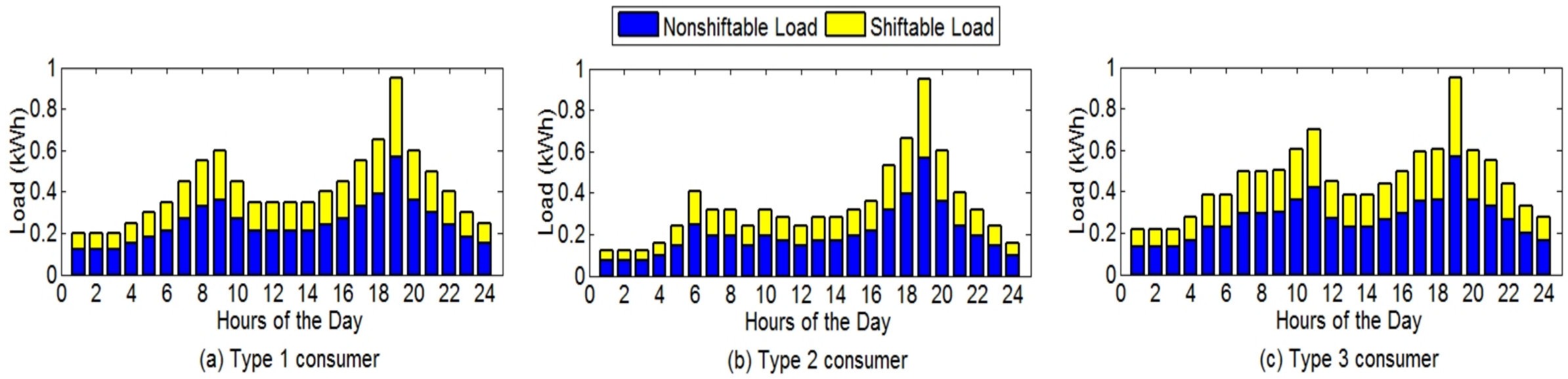

Fig.3. The Desired Power Consumption Patterns of Type 1, 2, 3 Consumers.

TABLE VI. COMPARISON OF TOTAL COSTS ACHIEVED BY DIFFERENT ALGORITHMS

| | | Number of consumers (Homogeneous, PAR<2.280) | | | | | Number of consumers (Heterogeneous, PAR<2.359 ) | | | | |
|---|---|---|---|---|---|---|---|---|---|---|---|
| | | 30 | 50 | 80 | 100 | 200 | 30 | 50 | 80 | 100 | 200 |
| Total cost | OG-DSM | 49.95 | 83.25 | 133.20 | 166.50 | 333.00 | 50.95 | 84.95 | 135.90 | 169.80 | 339.70 |
| | JO-DSM | 42.18 | 70.08 | 111.95 | 139.86 | 279.40 | 42.37 | 70.41 | 112.46 | 140.48 | 280.66 |
| | SC-DSM | 46.63 | 77.71 | 124.34 | 155.42 | 310.84 | 47.73 | 79.59 | 127.32 | 159.07 | 318.26 |
| | N-DSM | 30.78 | 50.78 | 80.78 | 100.78 | 200.78 | 26.23 | 43.25 | 68.35 | 84.50 | 168.73 |
| Performance gain | Over OG-DSM | 38% | 39% | 39% | 39% | 40% | 49% | 49% | 50% | 50% | 50% |
| | Over JO-DSM | 27% | 28% | 28% | 28% | 28% | 38% | 39% | 39% | 40% | 40% |
| | Over SC-DSM | 34% | 35% | 35% | 35% | 35% | 45% | 46% | 46% | 47% | 47% |

We divide the existing algorithms into three categories as shown in Table V.

The One-shot Game based stationary DSM (OG-DSM) algorithms with myopic price-anticipating consumers [6]-[8][15] calculate the NE and operate at NE of the one-shot energy scheduling game, which is characterized in Theorem 1. The Joint Optimization (JO-DSM) algorithms with myopic price-taking consumers [12][13] assume that the obedient consumers jointly minimize the total cost of the system. In this case, the optimal performance of stationary DSM mechanism can be achieved by appropriate pricing schemes. The Single-consumer Stochastic Control (SC-DSM) methods [9]-[11] try to use stochastic control methods to minimize the total cost of a single consumer. In this case, the utility company sets the price as $p_{\overline{h}} = p_{Hi}$ and $p_h = p_{Lo}, h \neq \overline{h}$, and the consumer buys energy in advance according to its scheduled power consumption pattern $a_i$. We assume that renewable energy is available with probability[5] $\varepsilon = 0.8$, in which case the consumer can reschedule its power consumption pattern to the desired pattern without suffering the discomfort cost since the energy supply is abundant. The renewable energy is not available with probability $1 - \varepsilon = 0.2$, in which case the consumer must comply with its scheduled power consumption pattern and will incur discomfort cost $d_i(a_i)$.

[5] This probability comes from the uncertainty of renewable energy generation (whether it is windy in wind energy generation, whether it is shiny in solar energy generation, etc.).

TABLE VII. COMPARISON WITH BILLING COST MINIMIZATION ALGORITHM

| Total Cost/ Performance gain | Number of consumers (Homogeneous) | | | | | Number of consumers (Heterogeneous) | | | | |
|---|---|---|---|---|---|---|---|---|---|---|
| | 30 | 50 | 80 | 100 | 200 | 30 | 50 | 80 | 100 | 200 |
| N-DSM | 36.21 | 60.09 | 95.52 | 119.40 | 238.80 | 35.21 | 58.39 | 92.82 | 116.10 | 232.10 |
| Billing cost minimization | 56.13 | 93.55 | 149.68 | 187.10 | 374.20 | 67.24 | 112.46 | 179.77 | 223.88 | 448.87 |
| Performance gain | 35% | 36% | 36% | 36% | 36% | 48% | 48% | 48% | 48% | 48% |

Given the same PAR goal, the comparison of total costs using these four algorithms is shown in Table VI. We can see that when the number of consumers increases, the N-DSM algorithm significantly outperforms other three algorithms. The cost reductions compared to OG-DSM, JO-DSM and SC-DSM are 40%, 28% and 35% in homogeneous case and 50%, 40% and 47% in heterogeneous case, respectively. Note that our algorithm, which is IC, can significantly outperform the JO-DSM algorithm, even though it is not IC.

### *B. Impact of Discomfort Cost*

Some works [6]-[8][16] consider minimizing the billing cost only, without taking into account the discomfort cost; while we consider the problem of jointly minimizing billing and discomfort costs as in [9]-[15]. We compare the results of the billing cost minimization algorithm with our proposed algorithm in Table VII. In the simulation, we assume that the PAR reduction goal is 10%. For the billing cost minimization algorithm, since there are numerous optimal solutions which achieve the minimal billing cost, we choose a fair solution where all consumers shift the same amount of peak-time consumption to off-peak times, and then calculate the billing and discomfort costs. By comparing the results in Table VII, we can see that the performance of our proposed algorithm significantly outperforms the billing cost minimization algorithm, with around 36% cost reduction for the homogeneous scenario and 48% cost reduction for the heterogeneous scenario, respectively. In fact, the billing cost minimization algorithm does not consider the impact of consumers' behavior on discomfort costs, resulting in a higher discomfort cost than our proposed N-DSM algorithm. The N-DSM algorithm induces the consumers to cooperate with each other to reduce their long-term discomfort costs and the consumers with higher discomfort costs benefit more through cooperation in the heterogeneous scenario. Thus, the performance gain of the N-DSM algorithm over the billing cost minimization policy in the heterogeneous scenario is higher than that in the homogeneous scenario.

*C. Impact of Consumer Preferences*

Recall from Section III that in order to achieve the social optimum, i.e., the solution to the DDP problem, the system runs the N-DSM algorithm which requires only a subset of the consumers to reduce their peak-time power usage. However, some consumers may have certain preferences of when to shift their consumption. Hence we simulate the scenario of $N = 3, m = 1$, to show the impact of consumer preferences. The consumers' types are Type 1, 2 and 3. We compare four cases:

- Case 1: all consumers can alter their power consumption patterns.
- Case 2: Type 1 consumer cannot alter its power consumption pattern on Mondays; Type 2 consumer cannot alter its power consumption pattern on Tuesdays; Type 3 consumer cannot alter its power consumption pattern on Wednesdays.
- Case 3: Type 1 and 3 consumers cannot alter their power consumption patterns on Mondays; Type 2 consumer cannot alter its power consumption pattern on Tuesdays.
- Case 4: Type 1, 2 and 3 consumers cannot alter their power consumption patterns on Mondays.

Note that the four cases represent different levels of heterogeneity in consumer preferences, which result in different levels of flexibility in the power consumption scheduling. In Case 1, no consumer has specific preferences on how to shift the power consumption pattern. Hence, the flexibility of scheduling is the highest. In Case 2, each consumer has distinct preferences on how to shift power consumption. Since their preferences are different, the flexibility of scheduling is still high. In Case 3, Type 1 and 3 consumers have similar preferences (i.e., both cannot shift on

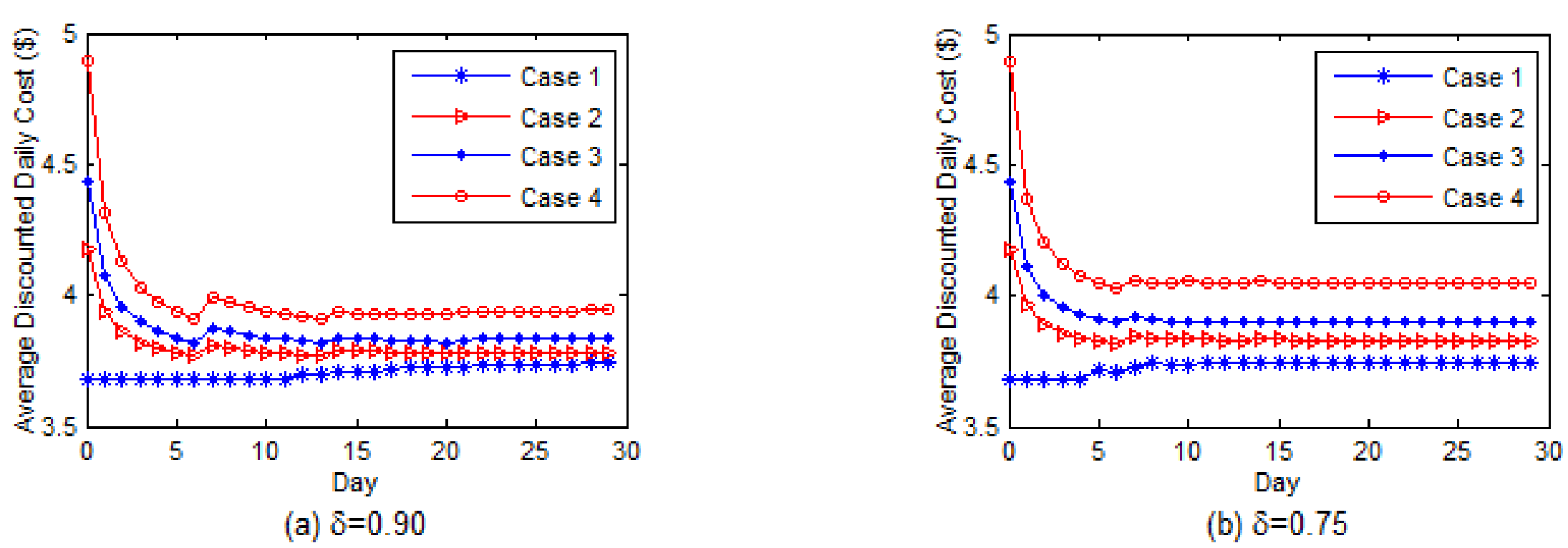

Fig.4. Comparison with Different Consumer Preferences.

TABLE VIII. IMPACT OF PERCENTAGE OF SHIFTABLE LOAD

| Total Cost | PAR Reduction goal | Percentage of shiftable load (Homogeneous) | | | | | Percentage of shiftable load (Heterogeneous) | | | | |
|---|---|---|---|---|---|---|---|---|---|---|---|
| | | 20% | 30% | 40% | 50% | 60% | 20% | 30% | 40% | 50% | 60% |
| JO-DSM | 10.0% | 166.80 | 155.40 | 149.30 | 145.80 | 143.20 | 186.80 | 168.80 | 159.10 | 153.60 | 149.50 |
| | 8.0% | 160.10 | 151.00 | 146.10 | 143.30 | 141.20 | 175.70 | 161.40 | 153.80 | 149.40 | 146.10 |
| | 6.7% | 155.60 | 148.00 | 144.00 | 141.60 | 139.90 | 168.80 | 156.80 | 150.40 | 146.80 | 144.00 |
| SC-DSM | 10.0% | -- | 161.70 | 155.40 | 149.20 | 142.90 | -- | 161.60 | 159.10 | 154.20 | 147.80 |
| | 8.0% | -- | 161.70 | 155.40 | 149.20 | 142.90 | -- | 161.60 | 159.10 | 154.20 | 147.80 |
| | 6.7% | -- | 161.70 | 155.40 | 149.20 | 142.90 | -- | 161.60 | 159.10 | 154.20 | 147.80 |
| N-DSM | 10.0% | 135.80 | 124.90 | 119.10 | 115.80 | 113.30 | 144.80 | 123.90 | 115.80 | 112.40 | 109.90 |
| | 8.0% | 128.60 | 120.00 | 115.30 | 112.60 | 110.60 | 130.20 | 116.70 | 112.00 | 109.30 | 107.30 |
| | 6.7% | 123.80 | 116.60 | 112.80 | 110.50 | 108.90 | 120.40 | 113.30 | 109.40 | 107.20 | 105.50 |

Mondays), which makes the scheduling less flexible (i.e. we can only alter Type 2 consumer's consumption on Mondays). In Case 4, all consumers have the same preference (i.e., all consumers cannot shift on Mondays). Hence, the scheduling is the least flexible.

We can see from Fig. 4 that the average discounted daily cost increases from Case 1 to Case 4, as expected. This implies that the system performance depends on the heterogeneity of consumer preferences. Regarding the impact of the discount factor on the convergence rate, we can see that there is no significant performance loss between the case when $\delta = 0.90$ and the case when $\delta = 0.75$.

*D. Impact of Percentage of Shiftable Load*

In order to evaluate the system performance with different percentages of the shiftable load, we show the total cost with different PAR reduction goals, as shown in Table VIII. The OG-DSM cannot meet the PAR reduction goals, so we compare the other three algorithms. We set the number of consumers to be $N = 100$. Simulation results show that given the constraint imposed by the PAR reduction goal, our algorithm can achieve around 8% and 9% reduction in the total cost when the percentage of shiftable load varies from 30% to 60%, in the homogeneous and heterogeneous scenarios, respectively; while these reductions are 7% and 10% with JO-DSM algorithm and 12% and 9% with SC-DSM algorithm. This implies that the performance of our algorithm is not significantly affected by consumers being homogeneous or heterogeneous, while the performances of the other two algorithms significantly depend on the consumer heterogeneity.

A trade-off between the optimal total cost and the PAR can be observed: when a higher threshold $l_{th}$ (i.e. a smaller $m$) is chosen, fewer consumers are required to change their power consumption patterns, resulting in a higher PAR but a lower discomfort cost. As a result, this trade-off should be considered when choosing the design parameter $l_{th}$ (or $m$).

TABLE IX. IMPACT OF THE LENGTH OF THE PEAK TIME

| Total Cost | | Number of consumers (Homogeneous) | | | | | Number of consumers (Heterogeneous) | | | | |
|---|---|---|---|---|---|---|---|---|---|---|---|
| | | 30 | 50 | 80 | 100 | 200 | 30 | 50 | 80 | 100 | 200 |
| JO-DSM | H=24 | 44.80 | 74.66 | 119.46 | 149.32 | 298.65 | 47.70 | 79.60 | 127.30 | 158.91 | 318.11 |
| | H=12 | 50.83 | 84.71 | 135.54 | 169.43 | 338.85 | 53.68 | 89.56 | 143.23 | 178.83 | 357.94 |
| | H=6 | 62.89 | 104.81 | 167.70 | 209.63 | 419.25 | 65.61 | 109.44 | 175.05 | 218.61 | 437.50 |
| SC-DSM | H=24 | 46.63 | 77.71 | 124.34 | 155.42 | 310.84 | 47.73 | 79.59 | 127.32 | 159.07 | 318.26 |
| | H=12 | 54.47 | 90.79 | 145.26 | 181.58 | 363.16 | 55.95 | 93.31 | 149.26 | 186.47 | 373.08 |
| | H=6 | 70.17 | 116.95 | 187.12 | 233.90 | 467.80 | 71.55 | 119.29 | 190.84 | 238.45 | 477.03 |
| N-DSM | H=24 | 36.21 | 60.09 | 95.52 | 119.40 | 238.80 | 35.60 | 59.31 | 93.78 | 116.62 | 234.31 |
| | H=12 | 36.59 | 60.71 | 96.48 | 120.60 | 241.20 | 35.59 | 59.01 | 93.78 | 117.30 | 234.50 |
| | H=6 | 37.36 | 61.96 | 98.40 | 123.00 | 246.00 | 36.36 | 60.26 | 95.70 | 119.70 | 239.30 |

*E. Impact of the Length of the Peak time*

In this subsection, we evaluate the system performance in terms of the total cost in Table IX, when the length of the peak time varies. We set the length of the peak time to be 1 hour ($H = 24$), 2 hours ($H = 12$) and 4 hours ($H = 6$). We need to change the desired power consumption pattern according to the length of the peak time (for example, the desired power consumption level in time slot 1 when $H = 12$ is the sum of the desired power consumption levels in time slots 1 and 2 when $H = 24$). We fix the PAR reduction goal at 10%. The OG-DSM cannot meet the PAR reduction goal. We compare the other three algorithms. First, we can see that our algorithm achieves a lower total cost than the other two under all lengths of the peak time with both homogeneous and heterogeneous consumers. Second, we can observe that the total cost rises when the length of the peak time increases. The increase in the total cost comes from the increased discomfort cost, because when the length of the peak time increases, the consumers need to shift more energy to off-peak times in order to obtain a low peak-time price. Hence, it is important to see how the performances of the algorithms vary with the length of the peak time. Under our algorithm, there is a slight increase in the total cost, i.e., 3% (homogeneous) and 2% (heterogeneous), respectively, when the length of the peak time varies from 1 hour to 4 hours. In contrast, the increases in the total cost are significant using the other two algorithms, namely 40% (homogeneous) and 38% (heterogeneous) for JO-DSM, and 50% (homogeneous) and 50% (heterogeneous) for SC-DSM. Clearly, our algorithm is less sensitive to the length of the peak time, yielding high performance for different lengths of the peak time as compared to the other two algorithms, whose performances degrade a lot when the length of the peak time increases.

## V. CONCLUSIONS

In this paper, we proposed a nonstationary DSM mechanism, which exploits the repeated interactions of the consumers over time. We rigorously prove that the proposed DSM mechanism can achieve the social optimum in terms of the total cost, and outperform existing stationary DSM strategies. Moreover, the proposed mechanism is IC, meaning that each self-interested consumer voluntarily follows the power consumption patterns recommended by the optimal DSM mechanism. Simulation results validate our analytical results on the DSM mechanism design and demonstrate up to 50% performance gains compared with existing mechanisms, especially when there are a large number of heterogeneous consumers in the systems. In addition, compared to the existing mechanisms, the performance of the proposed mechanism is much less sensitive and thus much more robust to system parameters such as the consumer heterogeneity and the length of the peak time.

## APPENDIX

### *A. Proof of Theorem 1*

Given the other consumers' action $a_{-i}$, we calculate the best response of consumer $i$. We denote $l_{-i,h} = \sum_{j \in \mathcal{N}, j \neq i} a_{j,h}$.

(a) If $l_{-i,h} + \overline{a}_{i,h} < l_{th}$, the best response of consumer $i$ is obviously $\overline{a}_i$, since $p_{Lo} A_i + d_i(\overline{a}_i) = p_{Lo} A_i$.

(b) If $l_{-i,h} + \overline{a}_{i,h} > l_{th}$, then $\forall a_i' \in \mathcal{A}_i^H, a_i' \neq \overline{a}_i$, we have

$$\begin{aligned} & p_{Lo} A_i + (p_{Hi} - p_{Lo}) \overline{a}_{i,\bar{h}} + d_i(\overline{a}_i) < p_{Lo} A_i + [\textstyle\sum_{h=1}^{H} k_{i,h}(|\, a_{i,h}' - \overline{a}_{i,h} \,|) + \omega_i] \\ & \leq \textstyle\sum_{h=1}^{H} p_h a_{i,h}' + d_i(a_i') \end{aligned}, \qquad (12)$$

where the first inequality is due to (6).

Therefore, the unique NE is $a_i^* = \overline{a}_i$.□

*B. Proof of Theorem 2*

We first prove that N-DSM can achieve the operating points in $\mathcal{B}$. Given any subset $I$ of $m$ consumer, let $i \in I$ choose action $\tilde{a}_i$ and $i \notin I(t)$ choose action $\overline{a}_i$. Then $i$ 's cost is $C_i^\dagger[I] = \tilde{c}_i \mathbf{1}_{\{i \in I\}} + \overline{c}_i \mathbf{1}_{\{i \notin I\}}$. It is easy to see that the cost profile $(C_1^\dagger[I], C_2^\dagger[I], \cdots, C_N^\dagger[I])$ is in $\mathcal{B}$. Since the other cost profiles in $\mathcal{B}$ are convex combinations of $\{(C_1^\dagger[I], C_2^\dagger[I], \cdots, C_N^\dagger[I])\}_I$, all the cost profiles in $\mathcal{B}$ can be achieved [24, Lemma 3.7.1].

Next, we prove $\mathcal{B}$ to be the Pareto-optimal region by showing that $\forall \boldsymbol{a}, \sum_{i=1}^N (c_i(\boldsymbol{a}) - \tilde{c}_i) / (\overline{c}_i - \tilde{c}_i) \geq m$ and equality holds only by choosing the action profiles described above, which proves the second statement of the theorem. To show these, we analyze the solution of the following problem:

$$\underset{(C_1, C_2, \cdots, C_N) \in \mathcal{C}}{\text{minimize}} \quad \sum_{i=1}^N (C_i - \tilde{c}_i) / (\overline{c}_i - \tilde{c}_i) \tag{13}$$

(a) Suppose $p_{\overline{h}}(\boldsymbol{a}) = p_{Hi}$, then the optimal action is $\overline{\boldsymbol{a}}$ and the corresponding optimal cost vector is $\boldsymbol{c}(\boldsymbol{a}) = (C_{1,NE}, C_{2,NE}, \cdots, C_{N,NE})$ due to Theorem 1. However,

$$\begin{aligned} \sum_{i=1}^N (C_{i,NE} - \tilde{c}_i) / (\overline{c}_i - \tilde{c}_i) &= \sum_{i=1}^N (p_{Hi} - p_{Lo}) \overline{a}_{i,\overline{h}} / d_i(\tilde{a}_i) \\ &\geq (p_{Hi} - p_{Lo}) \sum_{i=1}^N \overline{a}_{i,\overline{h}} / \max_{i \in \mathcal{N}} \{d_i(\tilde{a}_i)\} = (p_{Hi} - p_{Lo}) \overline{l}_{\overline{h}} / \max_{i \in \mathcal{N}} \{d_i(\tilde{a}_i)\} > m \end{aligned} . \tag{14}$$

where the last inequality is due to (6).

(b) Suppose $p_{\overline{h}}(\boldsymbol{a}) = p_{Lo}$ and the optimal action is $\boldsymbol{a}$. Obviously, $\sum_{i=1}^N a_{i,\overline{h}} = l_{th}$, otherwise at least one consumer can reduce its discomfort cost by shifting peak-time power while keeping the total billing cost and other consumers' discomfort costs unchanged. Next, we show that either $a_{i,\overline{h}} = \overline{a}_{i,\overline{h}}$ or $a_{i,\overline{h}} = b_{i,\overline{h}}$.

Suppose for $i$ and $j$, we have $b_{i,\overline{h}} < a_{i,\overline{h}} < \overline{a}_{i,\overline{h}}$ and $b_{j,\overline{h}} < a_{j,\overline{h}} < \overline{a}_{j,\overline{h}}$. Without loss of generality, we assume $(k_{i,\overline{h}} + \min_{h \neq \overline{h}} k_{i,h}) / d_i(\tilde{a}_i) \leq (k_{j,\overline{h}} + \min_{h \neq \overline{h}} k_{j,h}) / d_j(\tilde{a}_j)$. Then we compare the costs of $i$ and $j$ using action $\boldsymbol{a}'$, where $a'_{i,\overline{h}} = a_{i,\overline{h}} - \triangle$, $a'_{i,h} = \underset{\hat{\boldsymbol{a}} \in \mathcal{A},\, \hat{a}_{i,\overline{h}} = a_{i,\overline{h}} - \triangle}{\arg\min}\ c_i(\hat{\boldsymbol{a}}), h \neq \overline{h}$, $a'_{j,\bar{h}} = a_{j,\bar{h}} - \triangle$ , $a'_{j,h} = \underset{\hat{\boldsymbol{a}} \in \mathcal{A},\, \hat{a}_{j,\bar{h}} = a_{j,\bar{h}} - \triangle}{\arg\min}\, c_j(\hat{\boldsymbol{a}}), h \neq \bar{h}$ , with $\triangle = \min\{a_{i,\bar{h}} - b_{i,\bar{h}},\ \bar{a}_{j,\bar{h}} - a_{j,\bar{h}}\}$ , and $a'_n = a_n, n \neq i, j$. Thus, we have:

$$\begin{aligned}
&\frac{(c_i(\boldsymbol{a}) - \tilde{c}_i)}{(\bar{c}_i - \tilde{c}_i)} + \frac{(c_j(\boldsymbol{a}) - \tilde{c}_j)}{(\bar{c}_j - \tilde{c}_j)} = \frac{d_i(a_i)}{d_i(\tilde{a}_i)} + \frac{d_j(a_j)}{d_j(\tilde{a}_j)} \\
&= [\omega_i + (k_{i,\bar{h}} + \min_{h \neq \bar{h}} k_{i,h})(\bar{a}_{i,\bar{h}} - a_{i,\bar{h}})] / d_i(\tilde{a}_i) + [\omega_j + (k_{j,\bar{h}} + \min_{h \neq \bar{h}} k_{j,h})(\bar{a}_{j,\bar{h}} - a_{j,\bar{h}})] / d_j(\tilde{a}_j) \\
&\geq [\omega_i + (k_{i,\bar{h}} + \min_{h \neq \bar{h}} k_{i,h})(\bar{a}_{i,\bar{h}} - (a_{i,\bar{h}} - \Delta))] / d_i(\tilde{a}_i) + [\omega_j + (k_{j,\bar{h}} + \min_{h \neq \bar{h}} k_{j,h})(\bar{a}_{j,\bar{h}} - (a_{j,\bar{h}} + \Delta))] / d_j(\tilde{a}_j) \\
&\geq \frac{(c_i(\boldsymbol{a}') - \tilde{c}_i)}{(\bar{c}_i - \tilde{c}_i)} + \frac{(c_j(\boldsymbol{a}') - \tilde{c}_j)}{(\bar{c}_j - \tilde{c}_j)}
\end{aligned} \quad . (15)$$

We notice that either $a'_i = \tilde{a}_i$ or $a'_j = \bar{a}_j$, i.e., either $c_i(\boldsymbol{a}') = \bar{c}_i$ or $c_j(\boldsymbol{a}') = \tilde{c}_i$. By repeating this procedure, we can finally get a subset $I$, where $i \in I$ chooses action $\tilde{a}_i$, and $i \notin I$ chooses action $\bar{a}_i$.

Therefore, the solution of problem (13) is in the form of $C_i^\dagger[I]$ and the Pareto-optimal region can be written as (8). In addition, any other action profile will result in $\sum_{i=1}^N (c_i(\boldsymbol{a}) - \tilde{c}_i) / (\bar{c}_i - \tilde{c}_i) > m$. In other words, the other cost profiles in the set $\mathcal{B}$ cannot be achieved by stationary strategies. □

*C. Proof of Theorem 3*

Based on N-DSM, in time period $t$, $c_i^t = \tilde{c}_i$, $i \in I(t)$ and $c_i^t = \bar{c}_i, i \notin I(t)$. Then,

$$C_i = (1-\delta) \sum_{t=0}^{\infty} \delta^t [\tilde{c}_i \mathbf{1}_{\{i \in I(t)\}} + \bar{c}_i \mathbf{1}_{\{i \notin I(t)\}}] \quad (16)$$

where $\mathbf{1}_{\{\bullet\}}$ is the indicator function. We denote $\boldsymbol{g} = (g_1, g_2, \cdots, g_N)$ , $g_i = \dfrac{C_i - \tilde{c}_i}{\bar{c}_i - \tilde{c}_i}$ , $\bar{g}_i = \dfrac{\bar{C}_i - \tilde{c}_i}{\bar{c}_i - \tilde{c}_i}$ and $\mathcal{G}_{\bar{\boldsymbol{C}}} = \{\boldsymbol{g} \mid g_i = (C_i - \tilde{c}_i) / (\bar{c}_i - \tilde{c}_i),\ \exists \boldsymbol{C} \in \mathcal{B}_{\bar{\boldsymbol{C}}}\}$ . Then we notice that $g_i = (1-\delta)\sum_{t=0}^{\infty} \delta^t \mathbf{1}_{\{i \in I(t)\}}$ and $\boldsymbol{C} \in \mathcal{B}_{\bar{\boldsymbol{C}}}$ is equivalent to $\boldsymbol{g} \in \mathcal{G}_{\bar{\boldsymbol{C}}}$.

We denote by $g_i(t) = (1-\delta)\sum_{\tau=t}^{\infty} \delta^\tau \mathbf{1}_{\{i \in I(\tau)\}}$ and $g_i(0) = g_i$ the continuation cost at time $t$ and 0, respectively. Then we use a backward induction method to show that continuation cost at any given time $t$ can be decomposed of the current cost and the continuation cost at time $t+1$. We call the vector $\boldsymbol{g}(t) \in \mathcal{G}_{\bar{C}}$ a feasible vector and through the decomposition we will show that when the discount factor $\delta$ satisfies (11), the continuation cost at any time $t+1$ is also a feasible vector, namely $\boldsymbol{g}(t+1) \in \mathcal{G}_{\bar{C}}$. Since the original $\boldsymbol{g}(0) \in \mathcal{G}_{\bar{C}}$, we use mathematical induction to show this.

For time period $t$, suppose $\boldsymbol{g}(t) \in \mathcal{G}_{\bar{C}}$, then

$$g_i(t) = (1-\delta)\mathbf{1}_{\{i \in I(t)\}} + \delta g_i(t+1). \tag{17}$$

Or equivalently,

$$g_i(t+1) = \begin{cases} [g_i(t) - (1-\delta)] / \delta, & i \in I(t) \\ g_i(t) / \delta, & i \notin I(t) \end{cases}. \tag{18}$$

We need to show $\sum_{i=1}^{N} g_i(t+1) = m$ and $0 \le g_i(t+1) \le \bar{g}_i$. The former one is obvious due to hypothesis. For the latter one, $0 \le g_i(t+1) \le \bar{g}_i$ is obvious for $i \notin I(t)$, and for $i \in I(t)$, we also need

$$0 \le \frac{g_i(t) - (1-\delta)}{\delta} \le \bar{g}_i, i \in I(t). \tag{19}$$

This can be simplified as

$$\begin{cases} \delta \le \dfrac{1 - g_i(t)}{1 - \bar{g}_i} \\ \delta \ge 1 - g_i(t) \end{cases}, \ i \in I(t). \tag{20}$$

The first term is satisfied due to hypothesis. The second term requires

$$\delta \ge \max_{g(t) \in \mathcal{G}_{\bar{C}}} \min_{I(t) \subseteq \mathcal{I}} \max_{i \in I(t)} \{1 - g_i(t)\}. \tag{21}$$

Without loss of generality, we sort $g_i(t)$ in an decreasing order, namely,

$$g_1(t) \ge g_2(t) \ge \cdots \ge g_m(t) \ge \cdots \ge g_N(t). \tag{22}$$

It is easy to calculate:

$$\max_{i \in I(t)} \{1 - g_i(t)\} = 1 - g_{\max_{i \in I(t)} \{i\}}(t), \tag{23}$$

and

$$\min_{I(t) \in \mathcal{I}} \max_{i \in I(t)} \{1 - g_i(t)\} = (1 - g_m(t)). \tag{24}$$

Thus, the worst case $\boldsymbol{g}(t)$ in (21) is

$$\boldsymbol{g}(t)=(\overline{g}_1,\cdots,\overline{g}_{m-1},\frac{(m-\sum_{i=1}^{m-1}\overline{g}_i)}{(N-m+1)},\cdots,\frac{(m-\sum_{i=1}^{m-1}\overline{g}_i)}{(N-m+1)}). \tag{25}$$

Thus we require

$$\delta \geq 1-\frac{(m-\sum_{i=1}^{m-1}\overline{g}_i)}{(N-m+1)} . \tag{26}$$

Since $(m-\sum_{i=1}^{m-1}\overline{g}_i)\geq m-(m-1)$, $\mathcal{B}_{\overline{C}}$ is achievable, when (11) is satisfied.

*D. Proof of Theorem 4*

Obviously, $\boldsymbol{C}^*=(C_1^*,C_2^*,\cdots C_N^*)$ is in $\mathcal{B}_{\overline{C}}$. According to the proof of Theorem 3, $\boldsymbol{C}^*$ can be decomposed of current cost plus continuation cost:

$$C_i(\pi^*|_{\eta^t})=\tilde{c}_i+(\overline{c}_i-\tilde{c}_i)g_i(t)=(1-\delta)[\overline{c}_i\mathbf{1}_{\{i\in I(t)\}}+\tilde{c}_i\mathbf{1}_{\{i\notin I(t)\}}]+\delta C_i(\pi^*|_{\eta^{t+1}}). \tag{27}$$

When the discount factor satisfies (11), the optimal operating point $\boldsymbol{C}^*$ can be achieved by decomposing into current cost plus continuation cost in each period. We also notice that the best strategy of $I(t)$ in (21) is to choose $I(t)=\{1,2,\cdots m\}$, i.e., the indexed with the $m$ largest $g_i(t)$. Therefore, the proposed N-DSM can achieve optimal performance $\boldsymbol{C}^*$ in the system.

To show N-DSM is IC, we need to show that given the operating point $\boldsymbol{C}^*=(C_1^*,C_2^*,\cdots C_N^*)\in\mathcal{B}_{\overline{C}}$ and an arbitrary time period $t$, the continuation cost is the minimum cost achievable for consumer $i$. By decomposition, we have

$$C_i(\boldsymbol{\pi}^*|_{\eta^t})=(1-\delta)[\overline{c}_i\mathbf{1}_{\{i\in I(t)\}}+\tilde{c}_i\mathbf{1}_{\{i\notin I(t)\}}]+\delta C_i(\boldsymbol{\pi}^*|_{\eta^{t+1}}). \tag{28}$$

Obviously for consumer $i\notin I(t)$, it has no incentive to deviate from strategy $\boldsymbol{\pi}^*$ in period $t$. For consumer $i\in I(t)$, if it deviates from strategy $\boldsymbol{\pi}^*$, the minimum cost it can achieve is $c_i(\overline{a}_i,\boldsymbol{\pi}_{-i}^*(\eta^t))$, since the best response is to choose $\overline{a}_i$ according to Theorem 1. According to one-shot deviation principle [24], when consumer $i$ deviates, the other consumers will play NE strategy from time period $t+1$. So the cost from period $t$ is

$$C_i(\boldsymbol{\pi}'|_{\eta^t})=(1-\delta)c_i(\overline{a}_i,\boldsymbol{\pi}_{-i}^*(\eta^t))+\delta C_i(\boldsymbol{\pi}_{NE}|_{\eta^{t+1}})=C_{i,NE}. \tag{29}$$

The second equality holds because $c_i(\overline{a}_i, \boldsymbol{\pi}^*_{-i}(\eta^t)) = C_i(\boldsymbol{\pi}_{NE}|_{\eta^{t+1}}) = C_{i,NE}$. Since for $\boldsymbol{\pi}^*$, $C_i(\boldsymbol{\pi}^*|_{\eta^t}) \le \min\{C_{i,\max}, C_{i,NE}\} \le C_i(\boldsymbol{\pi}'|_{\eta^t})$, thus the consumers have no incentive to deviate from strategy $\boldsymbol{\pi}^*$ individually. Therefore, N-DSM is IC. □